\documentclass[aps, prb, twocolumn, showpacs, superscriptaddress]{revtex4-1}

\usepackage{amsmath,amssymb}
\usepackage{graphicx}
\usepackage{dcolumn}
\usepackage{bm}
\begin{document}

\title {One-dimensional Hubbard model with pair-hopping term: Pair-density wave, pair-superconductor and orthogonal metal}
\author{Yin Zhong}
\email{zhongy05@hotmail.com}
\affiliation{Center for Interdisciplinary Studies $\&$ Key Laboratory for
Magnetism and Magnetic Materials of the MoE, Lanzhou University, Lanzhou 730000, China}
\author{Lan Zhang}
\affiliation{Center for Interdisciplinary Studies $\&$ Key Laboratory for
Magnetism and Magnetic Materials of the MoE, Lanzhou University, Lanzhou 730000, China}
\author{Han-Tao Lu}
\affiliation{Center for Interdisciplinary Studies $\&$ Key Laboratory for
Magnetism and Magnetic Materials of the MoE, Lanzhou University, Lanzhou 730000, China}
\author{Hong-Gang Luo}
\email{luohg@lzu.edu.cn}
\affiliation{Center for Interdisciplinary Studies $\&$ Key Laboratory for
Magnetism and Magnetic Materials of the MoE, Lanzhou University, Lanzhou 730000, China}
\affiliation{Beijing Computational Science Research Center, Beijing 100084, China}

\begin{abstract}
As the simplest non-Fermi liquids, orthogonal metals have gapped single-particle excitations but show normal metallic behaviors in thermodynamic and transport quantities. Such an exotic metallic state could be realized in a Hubbard model with pair-hopping term. However, it is difficult to solve this model in higher dimension, here we focus on its one-dimension version, in which a powerful U(1) abelian bosonization technique is available to capture \textit{analytically} its universal low-energy physics. It is found that the one-dimensional system is unable to provide an appropriate precursor for the desirable orthogonal metals in higher dimension and the occurrence of the featureless orthogonal metallic phase needs \textit{at least} a quasi-one-dimensional multi-leg ladder model. Nevertheless, two interesting quantum liquid phases are identified for the present one-dimensional model, namely, pair-density wave (PDW) state and pair-superconducting (PSC) state. The former describes the charge-density-wave order of cooper pairs while the latter shows the superfluidity of two cooper pairs. Needing not any explicit calculation, these two phases could have further implications in the two-dimensional case, namely, the PDW state may give rise to quantum electron liquid crystal states, which is believed to be responsible for the complex pseudogap phase in cuprate, and the PSC state with charge 4e can be related to certain nematic superconducting state. These predications call for further investigation on this model in higher dimension by available analytical methods and/or numerical calculations.
\end{abstract}

\date{\today}

\maketitle
\section{Introduction}
Understanding the non-Fermi liquid behaviors,\cite{Sachdev2011} which exist ubiquitously in high-temperature cuprate superconductors,\cite{Lee2006,Armitage2010,Scalapino2012} iron-based superconductors and heavy fermion compounds,\cite{Chubukov2012,Steglich2005,Rosch,Stewart2011} is one of essential issues in condensed matter community.
However, because the cornerstone of interacting fermionic quantum liquid, namely, Landau Fermi liquid theory, seems to break down seriously,\cite{Sachdev2011} the secret of those exotic quantum liquid phases is still elusive and controversial. Recently, an important step with a simple picture toward the understanding of non-Fermi liquid appears. A kind of non-Fermi liquid named orthogonal metal, which has similar thermodynamic and transport properties of the classic Landau Fermi liquid but with gapped single-particle excitation and non-trivial topological feature, has been proposed based on the study of $Z_{2}$ slave-spin representation of electronic lattice Hamiltonian.\cite{Nandkishore,Ruegg2012} Many interesting extensions have been put forward and analyzed in detail with quantum field theory or exactly soluble models.\cite{Zhong2012a,Zhong2012b,Zhong2013,Maciejko2013,Maciejko2014,Zhong2014}

Irrespective of these intensive investigations, one should point out that the orthogonal metal has not been confirmed conclusively in any realistic lattice models though some authors have suggested a possible microscopic model to realize such an exotic quantum liquid state. The proposed model, \cite{Nandkishore} also called as Penson-Kolb-Hubbard model,\cite{Belkasri,Bossche,Dolcini2000,Dolcini2002,Dolcini2013} reads
\begin{eqnarray}
&& H=-t\sum_{\langle ij\rangle\sigma}(c_{i\sigma}^{\dag}c_{j\sigma}+H.c.)+U\sum_{i}c_{i\uparrow}^{\dag}c_{i\uparrow}c_{i\downarrow}^{\dag}c_{i\downarrow}\nonumber \\
&&\hspace{1.cm} -t'\sum_{\langle ij\rangle}(c_{i\uparrow}^{\dag}c_{i\downarrow}^{\dag}c_{j\downarrow}c_{j\uparrow}+H.c.),\label{eq0}
\end{eqnarray}
where $\langle \cdot \rangle$ denotes nearest-neighbor sites in the two-dimensional lattice. The first two terms correspond to the standard Hubbard model with $U$ being the onsite Coulomb energy between electrons with opposite spins and $t$ is the hopping energy. The third term is the pair-hopping term, which describes the hopping of a spin-singlet pair between nearest-neighbor site and can be produced by strong electron-phonon coupling.\cite{Assaad1997} It has been argued that such a pair-hopping term is able to induce the desirable orthogonal metal state if the condition $t' \gg U \gg t$ is fulfilled.\cite{Nandkishore}
Moreover, we should emphasize that the above Penson-Kolb-Hubbard model has been investigated by many analytical and numerical methods and they mainly focus on the global phase diagram and phase transitions.\cite{Belkasri,Bossche,Dolcini2000,Dolcini2002,Dolcini2013,Hui1993,Robaszkiewicz1999,Kapcia2012,Kapcia2013}
But, the mentioned possibility of orthogonal metal in such model motivates us to revisit this well-known model.

However, it is well-known that two-dimensional Hubbard-like models are still hard to solve by current analytical/numerical techniques due to non-perturbative nature, gigantic dimension of Hilbert space and minus-sign problem.  Fortunately, the one-dimensional version of Eq. (\ref{eq0}) can be attacked by many
powerful methods like Bethe ansatz,\cite{Andrei} abelian/non-abelian bosonization,\cite{Giamarchi,Gogolin} exact diagnolization and density-matrix-renormaliztion-group.\cite{White,Schollwock}

In the present work, we discuss the simplest but physically transparent abelian bosonization\cite{Giamarchi} treatment of one-dimensional Hubbard model with pair-hopping term. We use a powerful U(1) abelian bosonization technique, which can capture \textit{analytically} the universal low-energy physics of the model. It is found that two unexpected quantum liquid states, namely, pair-density wave state and pair-superconducting state, appear in our model calculation. The former describes the charge-density wave order of cooper pairs while the latter shows the superfluidity of two cooper pairs. It is noted that these two new phases are not reported in previous studies since generally they are the $4K_{F}$ singularity and are often neglected in many studies. However, neither PDW nor PSC corresponds to the wanted orthogonal metal and we suspect that no realistic one-dimensional lattice models can support the orthogonal metals.
Thus, we suggest that we should at least go to the quasi-one-dimensional multi-leg ladders models in order to get the featureless orthogonal metallic phase. When it comes to two-dimensional case, the PDW state may give rise to quantum electron liquid crystal states, which is believed to be responsible for the complex pseudogap phase in cuprate.\cite{Kivelson}

The remainder of the paper is organized as follows. In Sec. \ref{sec1}, the bosonization formalism of one-dimensional Hubbard model with a pair-hopping term is derived. In Sec. \ref{sec2}, the phase diagram is discussed. In Sec. \ref{sec3}, the relation to orthogonal metal and implication for high-dimension case are briefly analyzed. Finally, Sec.\ref{sec4} is devoted to a concise conclusion. We also present appendices $A$ - $E$ for the reader who is interested in the some details for the bosonization techniques.

\section{The model and its bosonization formalism} \label{sec1}
The model we used is the one-dimensional Hubbard model at generic filling supplemented with a pair-hopping term,
\begin{eqnarray}
&& H = -t\sum_{i\sigma}(c_{i\sigma}^{\dag}c_{i+1\sigma}+h.c.)+U\sum_{i}c_{i\uparrow}^{\dag}c_{i\uparrow}c_{i\downarrow}^{\dag}c_{i\downarrow}\nonumber \\
&&\hspace{1cm} -t'\sum_{i}(c_{i\uparrow}^{\dag}c_{i\downarrow}^{\dag}c_{i+1\downarrow}c_{i+1\uparrow}+h.c.) .\label{eq1}
\end{eqnarray}
It is clear that the first two terms consist the standard one-dimensional Hubbard model, which have the well-known bosonization formalism,\cite{Giamarchi}
namely, $H_{t}+H_{U}=H_{\rho}+H_{\sigma}$, where
\begin{eqnarray}
&&H_{\rho}=\frac{1}{2\pi}\int dx[u_{\rho}K_{\rho}(\partial_{x}\theta_{\rho})^{2}+\frac{u_{\rho}}{K_{\rho}}(\partial_{x}\phi_{\rho})^{2}]\nonumber\\
&&\hspace{1cm} -\frac{2U}{(2\pi\alpha)^{2}}\int dx\cos(2\sqrt{2}\phi_{\rho}-4k_{F}x),\label{eq2a}\\
&&H_{\sigma}=\frac{1}{2\pi}\int dx[u_{\sigma}K_{\sigma}(\partial_{x}\theta_{\sigma})^{2}+\frac{u_{\sigma}}{K_{\sigma}}(\partial_{x}\phi_{\sigma})^{2}]\nonumber\\
&&\hspace{1cm} +\frac{2U}{(2\pi\alpha)^{2}}\int dx\cos(2\sqrt{2}\phi_{\sigma}).\label{eq2b}
\end{eqnarray}
Here, the renormalized velocities $u_{\rho},u_{\sigma}$ and Luttinger parameters $K_{\rho},K_{\sigma}$ are defined as
$u_{\rho}K_{\rho}=u_{\sigma}K_{\sigma}=v_{F}$, $\frac{u_{\rho}}{K_{\rho}}=v_{F}+U/\pi$ and $\frac{u_{\sigma}}{K_{\sigma}}=v_{F}-U/\pi$. Thus, one can read the Luttinger parameter $K_{\rho}=1/\sqrt{1+U/(v_{F}\pi)}$ and $K_{\sigma}=1/\sqrt{1-U/(v_{F}\pi)}$. Generally, for positive (negative) U, $K_{\rho}<1$ ($K_{\rho}>1$) and $K_{\sigma}$ is fixed to unit if the spin part is gapless and the spin SU(2) symmetry is preserved. For the notations and formula used in this work one can refer to the book of Giamarchi.\cite{Giamarchi} In the following we focus on the bosonization of the pair-hopping term, which is of central interest in our work.

\subsection{Bosonization of the pair-hopping term}
The pair-hopping term
\begin{eqnarray}
&& H_{t'} = -t'\sum_{i}(c_{i\uparrow}^{\dag}c_{i\downarrow}^{\dag}c_{i+1\downarrow}c_{i+1\uparrow}+h.c.)\nonumber\\
&&\hspace{0.5cm} =-t'\int dx \psi_{\uparrow}^{\dag}(x)\psi_{\downarrow}^{\dag}(x)\psi_{\downarrow}(x+a)\psi_{\uparrow}(x+a)+h.c. \label{eq3}
\end{eqnarray}
will have many different terms and one has to do a tedious calculation when transforming it into its bosonized form.

The basic element is
\begin{eqnarray}
&&\psi_{\uparrow}^{\dag}(x)\psi_{\downarrow}^{\dag}(x)=\psi_{R\uparrow}^{\dag}\psi_{L\downarrow}^{\dag}+\psi_{L\uparrow}^{\dag}\psi_{R\downarrow}^{\dag}\nonumber\\
&&\hspace{1cm} +\psi_{R\uparrow}^{\dag}\psi_{R\downarrow}^{\dag}e^{-2ik_{F}x}+\psi_{L\uparrow}^{\dag}\psi_{L\downarrow}^{\dag}e^{+2ik_{F}x}\label{eq4}
\end{eqnarray}
and
\begin{eqnarray}
&& \psi_{\downarrow}(x+a)\psi_{\uparrow}(x+a) =\psi_{R\downarrow}\psi_{L\uparrow}+\psi_{L\downarrow}\psi_{R\uparrow}\nonumber\\
&&\hspace{1cm} +\psi_{R\downarrow}\psi_{R\uparrow}e^{-2ik_{F}(x+a)}+\psi_{L\downarrow}\psi_{L\uparrow}e^{+2ik_{F}(x+a)}.\label{eq5}
\end{eqnarray}
One should be aware that in Eq. (\ref{eq5}) each field in this equation is defined at $x+a$ instead of the ones in Eq. (\ref{eq4}).

Then, multiplying Eqs. (\ref{eq4}) and (\ref{eq5}), the resultant bosonized form reads [using Eq. (\ref{B2})]
\begin{eqnarray}
\frac{1}{(2\pi\alpha)^{2}}&&[e^{2ik_{F}a}e^{\sqrt{2}i(\phi_{\rho}-\phi_{\rho}'-\theta_{\rho}+\theta_{\rho}')}\nonumber\\
&&+e^{-2ik_{F}a}e^{\sqrt{2}i(-\phi_{\rho}+\phi_{\rho}'-\theta_{\rho}+\theta_{\rho}')}\nonumber\\
&&+e^{2ik_{F}a}e^{4ik_{F}x}e^{\sqrt{2}i(-2\phi_{\rho}-\theta_{\rho}+\theta_{\rho}')}\nonumber\\
&&+e^{-2ik_{F}a}e^{-4ik_{F}x}e^{\sqrt{2}i(2\phi_{\rho}-\theta_{\rho}+\theta_{\rho}')}\nonumber\\
&&+e^{\sqrt{2}i(2\phi_{\sigma}-\theta_{\rho}+\theta_{\rho}')}+e^{\sqrt{2}i(-2\phi_{\sigma}-\theta_{\rho}+\theta_{\rho}')}\nonumber\\
&&+e^{\sqrt{2}i(\phi_{\sigma}-\phi_{\sigma}'-\theta_{\rho}+\theta_{\rho}')}+e^{\sqrt{2}i(-\phi_{\sigma}+\phi_{\sigma}'-\theta_{\rho}+\theta_{\rho}')}],\nonumber
\end{eqnarray}
where the field $\phi',\theta'$ are defined at $x+a$. To proceed, we utilize $e^{\sqrt{2}i(\phi-\phi')}\simeq1-i\sqrt{2}a\partial_{x}\phi-a^{2}(\partial_{x}\phi)^{2}$ and $e^{\sqrt{2}i(\phi+\phi')}\simeq e^{2\sqrt{2}i\phi}$ (the case with $\theta$ is similar), we obtain
\begin{eqnarray}
\frac{1}{(2\pi\alpha)^{2}}&&[(-a^{2}\cos(2k_{F}a))(\partial_{x}\phi_{\rho})^{2}\nonumber\\
&&+(-a^{2}\cos(2k_{F}a)-a^{2})(\partial_{x}\theta_{\rho})^{2}\nonumber\\
&&-2a^{2}(\partial_{x}\phi_{\sigma})^{2}+4ia^{2}\sin(2k_{F}a)\partial_{x}\phi_{\rho}\partial_{x}\theta_{\rho}\nonumber\\
&&+2\cos(2\sqrt{2}\phi_{\sigma})+2\cos(2\sqrt{2}\phi_{\rho}-4k_{F}x-2k_{F}a)].\label{eq6}
\end{eqnarray}

With Eq. (\ref{eq6}) in hand, the full model Eq. (\ref{eq1}) can be rewritten as $H_{t}+H_{U}+H_{t'}=H_{\rho}+H_{\sigma}$, where
\begin{eqnarray}
&&H_{\rho}=\frac{1}{2\pi}\int dx[u_{\rho}K_{\rho}(\partial_{x}\theta_{\rho})^{2}+\frac{u_{\rho}}{K_{\rho}}(\partial_{x}\phi_{\rho})^{2}]\nonumber\\
&&\hspace{0.5cm} -\frac{2U}{(2\pi\alpha)^{2}}\int dx\cos(2\sqrt{2}\phi_{\rho}-4k_{F}x)\nonumber\\
&&\hspace{0.5cm} -\frac{4t'}{(2\pi\alpha)^{2}}\int dx\cos(2\sqrt{2}\phi_{\rho}-4k_{F}x-2k_{F}a),\nonumber\\
&&H_{\sigma}=\frac{1}{2\pi}\int dx[u_{\sigma}K_{\sigma}(\partial_{x}\theta_{\sigma})^{2}+\frac{u_{\sigma}}{K_{\sigma}}(\partial_{x}\phi_{\sigma})^{2}]\nonumber\\
&&\hspace{0.5cm} +\frac{2U-4t'}{(2\pi\alpha)^{2}}\int dx\cos(2\sqrt{2}\phi_{\sigma}).\label{eq7}
\end{eqnarray}

If we are interested in the generic fillings, the cosine term of charge part can be neglected and
$H_\rho$ can be further simplified as
\begin{equation}
H_{\rho}=\frac{1}{2\pi}\int dx[u_{\rho}K_{\rho}(\partial_{x}\theta_{\rho})^{2}+\frac{u_{\rho}}{K_{\rho}}(\partial_{x}\phi_{\rho})^{2}].\label{eq8}
\end{equation}
In the above expressions, we define $u_{\rho}K_{\rho}=v_{F}+2t'(1+\cos(2k_{F}a))/\pi,u_{\sigma}K_{\sigma}=v_{F}$, $\frac{u_{\rho}}{K_{\rho}}=v_{F}+U/\pi+2t'\cos(2k_{F}a)/\pi$ and $\frac{u_{\sigma}}{K_{\sigma}}=v_{F}-U/\pi+2t'/\pi$.
The Luttinger parameters become
$K_{\rho}=\sqrt{\frac{1+\frac{2t'(1+\cos(2k_{F}a))}{\pi v_{F}}}{1+\frac{U}{\pi v_{F}}+\frac{2t'\cos(2k_{F}a)}{\pi v_{F}}}}$ and $K_{\sigma}=\frac{1}{\sqrt{1-\frac{U}{\pi v_{F}}+\frac{2t'}{\pi v_{F}}}}$.

\section{Phase diagram of Hubbard model with pair-hopping term} \label{sec2}
In this section, we will analyze the structure of the phase diagram of Eqs. (\ref{eq7}) and (\ref{eq8}), which are valid at generic filling. For half-filling case it is well-known that the charge excitation is gapped, as in the usual one-dimensional Hubbard model.

The charge part is always gapless since the cosine term is absent. For the spin part, if $U-2t'>0$, the cosine term is marginally irrelevant and the spin part is also gapless with its Luttinger parameter $K_{\sigma}$ fixed to one.
However, if $U-2t'<0$ is satisfied, the spin part is gapped due to the cosine term and it generally belongs to the universal class of Luther-Emery liquid. At the same time, any correlation functions possessing $\theta_{\sigma}$ should show exponential decay. In the following we discuss them in detail.

\subsection{$U-2t'<0$}
In this case, the spin part is gapped and the surviving correlations are charge-density wave (CDW) or spin-singlet pairing (SSP). As it is known from standard bosonization result (see, e.g., Appendix E), CDW dominates when $K_{\rho}<1$ while SSP is the most important correlation for $K_{\rho}>1$.
However, for the present model with pair-hopping term, the pair-hopping correlation may be crucial. The pair-hopping order parameter is defined as  $O_{PH}(x)=\psi_{\uparrow}^{\dag}(x)\psi_{\downarrow}^{\dag}(x)\psi_{\downarrow}(x+a)\psi_{\uparrow}(x+a)\sim\cos(2\sqrt{2}\phi_{\sigma})+\cos(2\sqrt{2}\phi_{\rho}-4k_{F}x-2k_{F}a)$ and its correlation function is
\begin{eqnarray}
\langle O_{PH}(x)O_{PH}^{\dag}(0)\rangle\sim A+B\frac{e^{i4k_{F}x}}{x^{4K_{\rho}}}\nonumber
\end{eqnarray}
where A is a non-zero constant coming from the pinned $\phi_{\sigma}$ when $U-2t'<0$ while $B$ is a non-universal constant. To get more insight into this new order, we may define the cooper pair field $\psi_{B}^{\dag}(x)=\psi_{\uparrow}^{\dag}(x)\psi_{\downarrow}^{\dag}(x)$ and $\psi_{B}(x+a)=\psi_{\downarrow}(x+a)\psi_{\uparrow}(x+a)$. Then, $O_{PH}(x)=\psi_{B}^{\dag}(x)\psi_{B}(x+a)$ behaves like the density of cooper pairs and $\langle O_{PH}(x)O_{PH}^{\dag}(0)\rangle=\langle \psi_{B}^{\dag}(x)\psi_{B}(x+a) \psi_{B}^{\dag}(a)\psi_{B}(0)\rangle$ may be considered as the correlation of density of cooper pairs or CDW of cooper pair. Since such correlation is long-ranged, the CDW of cooper pairs
is a real order in this state and we may call it as pair-density wave (PDW). It is also clear that in contrast to the usual CDW which shows the fluctuation of charge e in real-space, the composite charge 2e cooper pair develops density wave order in the mentioned PDW.

So, we conclude that the $U-2t'<0$ case is dominated by PDW, which shows the behavior of CDW of cooper pairs. It should be emphasized that the PDW here is different from the one in Ref. [\onlinecite{Berg2010}], where their PDW is defined as $O_{PDW}(x)=(\psi_{\uparrow}^{\dag}(x)\psi_{\downarrow}^{\dag}(x+a)-\psi_{\downarrow}^{\dag}(x)\psi_{\uparrow}^{\dag}(x+a))$.

\subsection{$U-2t'>0$}
\begin{figure}
\includegraphics[width=6cm]{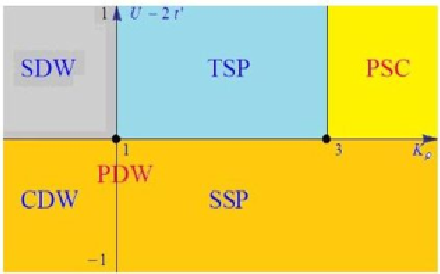}
\caption{The phase diagram for the Hubbard model with pair-hopping term. For $U-2t'>0$. SDW, TSP and PSC represent spin-density wave state, triplet superconducting state and pair-superconducting state, respectively, while for $U-2t'<0$, the pair-density wave (PDW), which shows the behavior of CDW of cooper pairs, is dominated in all regime. }
\label{fig:1}
\end{figure}

For $U-2t'>0$, the spin is gapless and one may suspect that the phase diagram in this case is identical to the case of standard Hubbard model whose phase diagram is discussed in detail in Appendix E.
However, we find that the pair-superconducting (PSC) correlation may be important when $K_{\rho}$ is large than 3.
This correlation is defined by $\langle O_{PS}(x)O_{PS}^{\dag}(0)\rangle$ with $O_{PS}(x)=\psi_{\uparrow}^{\dag}(x)\psi_{\downarrow}^{\dag}(x)\psi_{\uparrow}^{\dag}(x+a)\psi_{\downarrow}^{\dag}(x+a)
=\psi_{B}^{\dag}(x)\psi_{B}^{\dag}(x+a)\sim e^{-2\sqrt{2}i\theta_{\rho}}$. Thus, we have $\langle O_{PS}(x)O_{PS}^{\dag}(0)\rangle\sim\frac{1}{x^{4/K_{\rho}}}$, which means that the pair of two cooper pairs may develop quasi-long-range order. Meanwhile, spin-density wave (SDW) and spin-triplet pairing (TSP) correlations behave as $\langle O_{SDW}(x)O_{SDW}^{\dag}(0)\rangle\sim\frac{1}{x^{1+K_{\rho}}}$ and $\langle O_{TS}(x)O_{TS}^{\dag}(0)\rangle\sim\frac{1}{x^{1+1/K_{\rho}}}$.
Comparing these correlation functions, we can establish the phase diagram in Fig. \ref{fig:1}. We should remind the reader that the PSC state can be seen as a charge 4e superconducting state rather than the usual charge 2e superconductor.\cite{Berg2009} Furthermore, in two dimension, such charge 4e superconducting state is able to have non-trivial space structure and a superconducting liquid crystal phases is expected.\cite{Barci2009}

\subsection{The single particle excitation}
For the single particle excitation, in the case of $U-2t'<0$, it is gapped. In contrast, for $U-2t'>0$, the single particle correlation function reads (see Appendix D for the detailed calculation)
\begin{eqnarray}
\langle\psi_{R\alpha}(x)\psi^{\dag}_{R\alpha}(0)\rangle\sim\frac{1}{x^{\frac{K_{\rho}+1/K_{\rho}+K_{\sigma}+1/K_{\sigma}}{4}}}\sim\frac{1}{x^{\frac{K_{\rho}+1/K_{\rho}+2}{4}}}. \nonumber
\end{eqnarray}

The corresponding local density of states, which is able to be readily measured by local differential conductance in scanning tunneling microscopy, reads
\begin{eqnarray}
N(\omega)\sim\int dx \langle\psi_{R\alpha}(x)\psi^{\dag}_{R\alpha}(0)\rangle e^{ikx}|_{k\rightarrow\omega}\sim\omega^{\frac{K_{\rho}+1/K_{\rho}-2}{4}}, \nonumber
\end{eqnarray}
which shows the expected power-law behavior. Meanwhile, the particle momentum distribution function reads $n_{\uparrow}(k)\sim\int dxe^{-ikx}\langle\psi_{R\uparrow}(x)\psi^{\dag}_{R\uparrow}(0)\rangle|_{t=0}\sim|k-k_{F}|^{\frac{K_{\rho}+1/K_{\rho}-2}{4}}$, which shows the expected power-law behavior and suggests that the Luttinger liquid framework remains valid in spite of the introduction of brute pair-hopping term.

\section{Relation to orthogonal metal and implication for high-dimension cases} \label{sec3}
\subsection{Relation to orthogonal metal}
In the main text, we have studied the bosonization solution of the one-dimensional Hubbard with pair-hopping term. It is found that new orders, i.e., PDW and PSC emerge with the help of pair-hopping.
Obviously, in one dimension, neither PDW nor PSC corresponds to the wanted orthogonal metal state which by definition has no
overlap with Fermi liquid ground-state but has similar thermodynamics and transport properties of Fermi liquid. Strictly speaking, if there exists the orthogonal metal in one dimension and one considers Luttinger liquid replacing the role of Fermi liquid, the one-dimensional orthogonal metal should have gapped single-particle excitation but with similar collective behavior (spin-charge separation) to the Luttinger liquid. From our calculation and analysis in last section, we have to conclude that no such counterpart of orthogonal metal exists in our present model and we think this conclusion is not limited by our bosonization treatment since such technique is believed to faithfully capture low energy physics of generic one-dimensional models.

\subsection{Implication for higher-dimension case}
Since no orthogonal metals are found in one dimension from our previous discussion, we have to move to the two-dimensional case.
To proceed, we assume that the bosonization results can be used at least qualitatively to understand what happens in two dimension. Firstly, the pair-density wave state may become as a macroscopic ordered state with a spin gap, which should be observed in both single-particle excitation spectrum and spin-spin correlation function. In contrast, the charge dynamics is free of gap and we expect good metallic behaviors for charge transport. More exciting
case appears when the pair-density wave is not uniform and such situation relates to various kinds of electronic nematic or sematic states where the translation or rotation symmetry breaks spontaneously.
The mentioned electron liquid crystal states may be responsible for the complex pseudogap phase in cuprate.\cite{Kivelson}

\section{Conclusion} \label{sec4}
We have studied the one-dimensional Hubbard model with pair-hopping term, which provides certain link to the orthogonal metals. We find that the one-dimensional system may be unable to provide an appropriate precursor for the desirable states. This suggests that a more detailed study on quasi-one-dimensional multi-leg ladders or higher dimensional cases are required. However, we identify two interesting quantum liquid phases, i.e. pair-density wave state and pair-superconducting state in 1D case. The former one describes the charge-density wave order of cooper pairs, whose high dimensional generalization may give rise to interesting quantum electron liquid crystal states. The latter pair-superconducting state presents the superfluidity of two cooper pairs, and a charge 4e nematic superconducting state is expected in two dimension subsequently.
We hope the present work may be helpful for further study in non-Fermi liquid and exotic pairing/supercodncuting states in strongly correlated systems.

\begin{acknowledgments}
The work was supported partly by NSFC, the Fundamental Research Funds for the Central Universities and the national program for basic research of China.
\end{acknowledgments}

\appendix
\section{Bosonization of the hopping term}
\begin{eqnarray}
H_{t}=-t\sum_{i\sigma}(c_{i\sigma}^{\dag}c_{i+1\sigma}+h.c.)=\sum_{k\sigma}-2t\cos(k)c_{k\sigma}^{\dag}c_{k\sigma}\label{A1}
\end{eqnarray}

The spectrum is $\varepsilon_{k}=-2t\cos(k)$ and the Fermi surface is determined by the external given Fermi energy $E_{F}$ as $\varepsilon_{k_{F}}=-2t\cos(k_{F})=E_{F}$.
Since the system is one-dimensional, the Fermi surface is in fact two separated points called Fermi points whose position located at $-k_{F}$ and $+k_{F}$.

If only low-energy physics is relevant, one can expand Eq. (\ref{A1}) around those two Fermi points ($\pm k_{F}$)
\begin{eqnarray}
&&\sum_{k\sigma}(-2t\cos(k)-E_{F})c_{k\sigma}^{\dag}c_{k\sigma}\nonumber\\
&&\hspace{1cm} \simeq \sum_{k\sigma}v_{F}k c_{k\sigma R}^{\dag}c_{k\sigma R}-v_{F}k c_{k\sigma L}^{\dag}c_{k\sigma L},\label{A2}
\end{eqnarray}
where we have defined right-hand field $c_{k\sigma R}$ (expanding around $+k_{F}$) and left-hand field $c_{k\sigma L}$ (expanding around $-k_{F}$). The Fermi velocity $v_{F}=2t\sin(k_{F})$  and the Fermi energy are explicitly introduced.
Then, transforming Eq.\ref{A2} into real space, we have
\begin{eqnarray}
H_{t}\simeq-iv_{F}\int dx[\psi_{R\sigma}^{\dag}\partial_{x}\psi_{R\sigma}-\psi_{L\sigma}^{\dag}\partial_{x}\psi_{L\sigma}],\label{A3}
\end{eqnarray}
where $\psi_{R},\psi_{L}$ are standard Fermi field defined in real space and we also have $\psi_{\sigma}(x)=\psi_{R\sigma}e^{ik_{F}x}+\psi_{L\sigma}e^{-ik_{F}x}$.

\section{Bosonization of free fermions with spin}
In this subsection, we present the bosonized formalism for free spinful fermions of Eq. (\ref{A3}).
\begin{eqnarray}
&&H_{t}\simeq-iv_{F}\int dx[\psi_{R\sigma}^{\dag}\partial_{x}\psi_{R\sigma}-\psi_{L\sigma}^{\dag}\partial_{x}\psi_{L\sigma}]\nonumber\\
&&=\frac{v_{F}}{2\pi}\int dx[(\partial_{x}\theta_{\uparrow})^{2}+(\partial_{x}\phi_{\uparrow})^{2}+(\partial_{x}\theta_{\downarrow})^{2}+(\partial_{x}\phi_{\downarrow})^{2}]\nonumber\\
&&=\frac{v_{F}}{2\pi}\int dx[(\partial_{x}\theta_{\rho})^{2}+(\partial_{x}\phi_{\rho})^{2}+(\partial_{x}\theta_{\sigma})^{2}+(\partial_{x}\phi_{\sigma})^{2}]\label{B1}
\end{eqnarray}
\begin{eqnarray}
&&\psi_{R\alpha}=\frac{1}{\sqrt{2\pi \alpha}}e^{i(\phi_{\alpha}-\theta_{\alpha})}=\frac{1}{\sqrt{2\pi \alpha}}e^{\frac{i}{\sqrt{2}}(\phi_{\rho}-\theta_{\rho}+\alpha(\phi_{\sigma}-\theta_{\sigma}))};\nonumber\\
&&\psi_{L\alpha}=\frac{1}{\sqrt{2\pi \alpha}}e^{i(-\phi_{\alpha}-\theta_{\alpha})}=\frac{1}{\sqrt{2\pi \alpha}}e^{\frac{i}{\sqrt{2}}(-\phi_{\rho}-\theta_{\rho}+\alpha(-\phi_{\sigma}-\theta_{\sigma}))};\nonumber\\
&&\phi_{\rho}=\frac{1}{\sqrt{2}}(\phi_{\uparrow}+\phi_{\downarrow});\phi_{\sigma}=\frac{1}{\sqrt{2}}(\phi_{\uparrow}-\phi_{\downarrow});\nonumber\\
&&\theta_{\rho}=\frac{1}{\sqrt{2}}(\theta_{\uparrow}+\theta_{\downarrow});\theta_{\sigma}=\frac{1}{\sqrt{2}}(\theta_{\uparrow}-\theta_{\downarrow});\nonumber\\
&&\psi_{r\alpha}^{\dag}\psi_{r\alpha}=\rho_{r\alpha};\rho_{R\alpha}+\rho_{L\alpha}=\rho_{\alpha};\nonumber\\
&&\frac{1}{\sqrt{2}}(\rho_{\uparrow}+\rho_{\downarrow})=\rho_{\rho}=-\frac{1}{\pi}\partial_{x}\phi_{\rho};\frac{1}{\sqrt{2}}(\rho_{\uparrow}-\rho_{\downarrow})=\rho_{\sigma}=\frac{1}{\pi}\partial_{x}\phi_{\sigma};\nonumber\\
&&\psi_{\alpha}^{\dag}(x)\psi_{\alpha}(x)=-\frac{1}{\pi}\partial_{x}\phi_{\alpha}+\frac{2}{2\pi\alpha}\cos(2\phi_{\alpha}-2k_{F}x)\nonumber\\
&&\label{B2}
\end{eqnarray}

Equation (\ref{A3}) now described the separated charge ($\theta_{\rho},\phi_{\rho}$) and spin ($\theta_{\sigma},\phi_{\sigma}$) degrees of freedom.

\section{Bosonization of the Hubbard-U term}
The Hubbard-U term is bosonized as follows
\begin{eqnarray}
H_{U}&&=U\sum_{i}c_{i\uparrow}^{\dag}c_{i\uparrow}c_{i\downarrow}^{\dag}c_{i\downarrow}\simeq U\int dx \psi_{\uparrow}^{\dag}(x)\psi_{\uparrow}(x)\psi_{\downarrow}^{\dag}(x)\psi_{\downarrow}(x)\nonumber\\
&&=U\int dx(-\frac{1}{\pi}\partial_{x}\phi_{\uparrow}+\frac{1}{2\pi\alpha}[e^{2ik_{F}x}e^{-2i\phi_{\uparrow}}+e^{-2ik_{F}x}e^{+2i\phi_{\uparrow}}])\nonumber\\
&&\hspace{1cm} \times(-\frac{1}{\pi}\partial_{x}\phi_{\downarrow}+\frac{1}{2\pi\alpha}[e^{2ik_{F}x}e^{-2i\phi_{\downarrow}}+e^{-2ik_{F}x}e^{+2i\phi_{\downarrow}}])\nonumber\\
&&=U\int dx[\frac{1}{\pi^{2}}\partial_{x}\phi_{\uparrow}\partial_{x}\phi_{\downarrow}+\frac{2}{(2\pi\alpha)^{2}}\cos(2(\phi_{\uparrow}-\phi_{\downarrow}))\nonumber\\
&&\hspace{1cm} -\frac{2}{(2\pi\alpha)^{2}}\cos(2(\phi_{\uparrow}+\phi_{\downarrow})-4k_{F}x)]\nonumber\\
&&=\frac{U}{2\pi^{2}}\int dx[(\partial_{x}\phi_{\rho})^{2}-(\partial_{x}\phi_{\sigma})^{2}]+\int dx\frac{2U}{(2\pi\alpha)^{2}}\cos(2\sqrt{2}\phi_{\sigma})\nonumber\\
&&\hspace{1cm} -\int dx\frac{2U}{(2\pi\alpha)^{2}}\cos(2\sqrt{2}\phi_{\rho}-4k_{F}x).\label{C1}
\end{eqnarray}

Here, the $4k_{F}$ term has a minus sign due to the Klein factor.[We can assign real fermion (Majorana fermion) $\eta_{R}$ and $\eta_{L}$ in front of $\psi_{R\alpha}$ and $\psi_{L\alpha}$ when four-fermion interaction is involved. Thus, the $4k_{F}$ term have factor $\eta_{R}\eta_{L}\eta_{R}\eta_{L}=-\eta_{R}\eta_{R}\eta_{L}\eta_{L}=-1$ while the cosine term of spin part has $\eta_{R}\eta_{L}\eta_{L}\eta_{R}=1$.]

Then, combining Eqs. (\ref{B2}) and (\ref{C1}), the standard Hubbard model is bosonized as
\begin{eqnarray}
&&H_{t}+H_{U}=H_{\rho}+H_{\sigma}.\nonumber\\
&&H_{\rho}=\frac{1}{2\pi}\int dx[u_{\rho}K_{\rho}(\partial_{x}\theta_{\rho})^{2}+\frac{u_{\rho}}{K_{\rho}}(\partial_{x}\phi_{\rho})^{2}]\nonumber\\
&&\hspace{1cm}-\frac{2U}{(2\pi\alpha)^{2}}\int dx\cos(2\sqrt{2}\phi_{\rho}-4k_{F}x),\nonumber\\
&&H_{\sigma}=\frac{1}{2\pi}\int dx[u_{\sigma}K_{\sigma}(\partial_{x}\theta_{\sigma})^{2}+\frac{u_{\sigma}}{K_{\sigma}}(\partial_{x}\phi_{\sigma})^{2}]\nonumber\\
&&\hspace{1cm}+\frac{2U}{(2\pi\alpha)^{2}}\int dx\cos(2\sqrt{2}\phi_{\sigma}).\label{C2}
\end{eqnarray}
The renormalized velocity $u_{\rho},u_{\sigma}$ and Luttinger parameter $K_{\rho},K_{\sigma}$ are defined as
$u_{\rho}K_{\rho}=u_{\sigma}K_{\sigma}=v_{F}$, $\frac{u_{\rho}}{K_{\rho}}=v_{F}+U/\pi$ and $\frac{u_{\sigma}}{K_{\sigma}}=v_{F}-U/\pi$. Thus, one can read the Luttinger parameter
$K_{\rho}=1/\sqrt{1+U/(v_{F}\pi)}$ and $K_{\sigma}=1/\sqrt{1-U/(v_{F}\pi)}$. Generally, for positive (negative) U, $K_{\rho}<1$ ($K_{\rho}>1$) and $K_{\sigma}$ is fixed to unit if the spin part is gapless
and the spin SU(2) symmetry is preserved.

\section{Some useful integral results}
If the bosonic fields $\phi,\theta$ are described in terms of the free bosons Hamiltonian
\begin{eqnarray}
\frac{1}{2\pi}\int dx[uK(\partial_{x}\theta)^{2}+\frac{u}{K}(\partial_{x}\phi)^{2}],\label{D1}
\end{eqnarray}
then the corresponding correlations read as
\begin{eqnarray}
\langle [\phi(x)-\phi(0)]^{2}\rangle\simeq\frac{K}{2}\ln(x^{2})=\ln(x^{K})\label{D2}
\end{eqnarray}
and
\begin{eqnarray}
\langle [\theta(x)-\theta(0)]^{2}\rangle\simeq\frac{1}{2K}\ln(x^{2})=\ln(x^{1/K}).\label{D3}
\end{eqnarray}

Then, using these two results, we have the single particle Green's function for the right mover
\begin{eqnarray}
\langle \psi_{R}(x)\psi_{R}^{\dag}(0)\rangle &&=\frac{1}{2\pi\alpha}\langle e^{-i(\phi(x)-\phi(0))+i(\theta(x)-\theta(0))}\rangle \nonumber\\
&&\simeq\frac{1}{2\pi\alpha}\langle  e^{-i(\phi(x)-\phi(0))}\rangle \langle e^{i(\theta(x)-\theta(0))}\rangle\nonumber\\
&&=\frac{1}{2\pi\alpha}e^{-\frac{1}{2}\langle(\phi(x)-\phi(0))^{2}\rangle}e^{-\frac{1}{2}\langle(\theta(x)-\theta(0))^{2}\rangle}\nonumber\\
&&\simeq\frac{1}{2\pi\alpha}e^{-\frac{1}{2}\ln(x^{K})}e^{-\frac{1}{2}\ln(x^{1/K})}\nonumber\\
&&\sim \frac{1}{x^{\frac{K}{2}+\frac{1}{2K}}}.\label{D4}
\end{eqnarray}
Particles with spin or other degree of freedom have similar results and one can use the same method here to obtain them, for example,
\begin{eqnarray}
\langle \psi_{R\alpha}(x)\psi_{R\alpha}^{\dag}(0)\rangle &&=\frac{1}{2\pi\alpha}\langle e^{-i(\phi_{\alpha}(x)-\phi_{\alpha}(0))+i(\theta_{\alpha}(x)-\theta_{\alpha}(0))}\rangle \nonumber\\
&&\simeq\frac{1}{2\pi\alpha}\langle  e^{-i(\phi_{\alpha}(x)-\phi_{\alpha}(0))}\rangle \langle e^{i(\theta_{\alpha}(x)-\theta_{\alpha}(0))}\rangle\nonumber\\
&&=\frac{1}{2\pi\alpha}\langle  e^{-\frac{i}{\sqrt{2}}(\phi_{\rho}(x)-\phi_{\rho}(0))}\rangle\langle  e^{-\frac{i}{\sqrt{2}}(\alpha\phi_{\sigma}(x)-\alpha\phi_{\sigma}(0))}\rangle\nonumber\\
&&\times\langle  e^{-\frac{i}{\sqrt{2}}(\theta_{\rho}(x)-\theta_{\rho}(0))}\rangle\langle  e^{-\frac{i}{\sqrt{2}}(\alpha\theta_{\sigma}(x)-\alpha\theta_{\sigma}(0))}\rangle \nonumber\\
&&=\frac{1}{2\pi\alpha}e^{-\frac{1}{4}\langle(\phi_{\rho}(x)-\phi_{\rho}(0))^{2}\rangle}e^{-\frac{1}{4}\langle(\theta_{\rho}(x)-\theta_{\rho}(0))^{2}\rangle}\nonumber\\
&&\times e^{-\frac{1}{4}\langle(\phi_{\sigma}(x)-\phi_{\sigma}(0))^{2}\rangle}e^{-\frac{1}{4}\langle(\theta_{\sigma}(x)-\theta_{\sigma}(0))^{2}\rangle}\nonumber\\
&&\simeq e^{-\frac{1}{4}[\ln(x^{K_{\rho}}) + \ln(x^{\frac{1}{K_{\rho}}})]}e^{-\frac{1}{4}[\ln(x^{K_{\sigma}}) + \ln(x^{\frac{1}{K_{\sigma}}})]}\nonumber\\
&&= \frac{1}{x^{\frac{1}{4}(K_{\rho}+1/K_{\rho}+K_{\sigma}+1/K_{\sigma})}}.\label{D5}
\end{eqnarray}

\section{Phase diagram of the Hubbard model}
If we are interested in generic filling, the cosine term of $\phi_{\rho}$ is irrelevant and the charge part $H_{\rho}$ is gapless.
The spin part has the cosine term and if $U>0$ the cosine term is marginally irrelevant, thus the spin part is gapless. In contrast, when $U<0$,
the cosine term will drive the spin part into a gapped phase.

In the gapped phase for spin, the bosonic field seems to be pinned to $\phi_{\sigma}=0+\frac{2\pi n}{2\sqrt{2}}$, which maximizes the contribution of the cosin term. Expanding around those
classical manifold, one can see a gap $|U|$ for the spin part.

\subsubsection{$U>0$ case}
For the phase diagram, the $U>0$ case is gapless for both charge and spin part. We introduce the SDW correlation, which is defined as
$\langle O(x)_{SDW \alpha}^{\dag}O(0)_{SDW \alpha}\rangle$
with $O(x)_{SDW \alpha}=\psi_{\sigma}^{\dag}(x)\sigma^{\alpha}_{\sigma\sigma'}\psi_{\sigma'}(x)$. The bosonized form of the SDW parameter is
\begin{eqnarray}
&&O(x)_{SDW x}\sim e^{-2ik_{F}x}\cos(\sqrt{2}\phi_{\rho})\cos(\sqrt{2}\theta_{\sigma})\nonumber\\
&&O(x)_{SDW y}\sim e^{-2ik_{F}x}\cos(\sqrt{2}\phi_{\rho})\sin(\sqrt{2}\theta_{\sigma})\nonumber\\
&&O(x)_{SDW z}\sim e^{-2ik_{F}x}\cos(\sqrt{2}\phi_{\rho})\sin(\sqrt{2}\phi_{\sigma}).\nonumber
\end{eqnarray}
The SDW correlation is calculated as $\langle O(x)_{SDW \alpha}^{\dag}O(0)_{SDW \alpha}\rangle\sim \frac{1}{x^{1+K\rho}}$ since $K_{\sigma}$ is fixed to 1.
In fact, the CDW correlation $O(x)_{CDW}=\psi_{\sigma}^{\dag}(x)\psi_{\sigma}(x)\sim e^{-2ik_{F}x}\sin(\sqrt{2}\phi_{\rho})\cos(\sqrt{2}\phi_{\sigma})$ has similar power-law decay, but due to an extra logarithmic correction for SDW correlation, the SDW wins over CDW finally.
There are also singlet and triplet superconducting correlation $O(x)_{SS}=\sigma\psi_{\sigma}^{\dag}(x)\psi^{\dag}_{-\sigma}(x)\sim e^{-i\sqrt{2}\theta_{\rho}}\cos(\sqrt{2}\phi_{\sigma})$ and $O(x)_{TS \alpha}=\sigma\psi_{\sigma}^{\dag}(x)\sigma^{\alpha}_{\sigma\sigma'}\psi^{\dag}_{\sigma'}(x)$, whose bosonized formalism is
\begin{eqnarray}
&&O(x)_{TS x}\sim e^{-i\sqrt{2}\theta_{\rho}}\cos(\sqrt{2}\theta_{\sigma})\nonumber\\
&&O(x)_{TS y}\sim -e^{-i\sqrt{2}\theta_{\rho}}\sin(\sqrt{2}\theta_{\sigma})\nonumber\\
&&O(x)_{TS z}\sim e^{-i\sqrt{2}\theta_{\rho}}\sin(\sqrt{2}\phi_{\sigma}).\nonumber
\end{eqnarray}
Then, $\langle O(x)_{SS}^{\dag}O(0)_{SS}\rangle\sim \langle O(x)_{TS\alpha}^{\dag}O(0)_{TS\alpha}\rangle\sim \frac{1}{x^{1+1/K_{\rho}}}$.
Comparing the power-law decay of different correlation, we find the TS correlation dominates when $K_{\rho}>1$ and the SDW correlation wins if $K_{\rho}<1$.

\subsubsection{$U<0$ case}
If the $U<0$, the spin part is gapped and we find the CDW dominates if $K_{\rho}<1$ while SS is the dominate correlation when $K_{\rho}>1$.
The power-decay of these two ones are $\langle O(x)_{CDW}^{\dag}O(0)_{CDW}\rangle\sim \frac{1}{x^{K_{\rho}}}$, $\langle O(x)_{SS}^{\dag}O(0)_{SS}\rangle\sim \frac{1}{x^{1/K_{\rho}}}$.

\subsubsection{Hubbard model at half-filling}
When half-filling is considered, we have $k_{F}=\frac{\pi}{2a}$,[$k_{F}$ is assumed to be fixed by filling the free particle energy band and it has the same spirit ($k_{F}$ or Fermi surface is not disturbed by interaction.) as Landau Fermi liquid.] thus $4k_{F}x=4\frac{\pi}{2a}na=2n\pi$ and the cosine term in charge part is relevant now. Then, such relevant term leads to gap for charge excitation and we see the Mott insulator in one dimension.

\end{document}